# Some Applications envisaged for the new generation of communications networks 6G

1st S. Latreche
Université de Bejaia, Faculté de Technologie, Laboratoire de Génie Electrique, 06000 Bejaia, Algeria
sofiane.latreche@univ-bejaia.dz

2nd H. Bellahsene
Université de Bejaia, Faculté de Technologie, Laboratoire de Génie Electrique, 06000 Bejaia, Algeria
hocine.bellahsene@univ-bejaia.dz

3rd A. Taleb-Ahmed
Valenciennes University
IEMN DOAE Laboratory
abdelmalik.taleb-ahmed@uphf.fr

*Abstract*—Future applications such as intelligent vehicles, the Internet of Things and holographic telepresence are already highlighting the limits of existing fifth-generation (5G) mobile networks. These limitations relate to data throughput, latency, reliability, availability, processing, connection density and global coverage, whether terrestrial, submarine or space-based. To remedy this, research institutes have begun to look beyond IMT2020, and as a result, 6G should provide effective solutions to 5G's shortcomings. 6G will offer high quality of service and energy efficiency to meet the demands of future applications that are unimaginable to most people. In this article, we present the future applications and services promised by 6G.

## I. INTRODUCTION

The evolution of mobile networks has witnessed remarkable growth since their inception in the 1980s. Transitioning from one generation to the next occurred at an astonishing pace. Initially dedicated solely to telephony, these networks then gradually integrated Short Message Service (SMS) and Multimedia Messaging Service (MMS). Subsequently, they evolved into a more multimedia-oriented generation with 3G technology, followed by a focus on high-speed data transmission with the advent of 4G and the establishment of all-IP networks. The introduction of 5G technology further pushed the boundaries, reducing latency and providing exceptionally high-speed, high-capacity connections that rendered services and applications previously deemed impossible just two decades ago not only feasible but also seamless [1].

However, the rapid development of future applications, such as smart vehicles, the Internet of Things (IoT), and holographic telepresence, is already revealing the constraints of existing 5th generation (5G) mobile networks. These constraints encompass aspects like data transmission rates, latency, reliability, network availability, processing capabilities, connection density, and global coverage, whether on terrestrial, submarine, or space-based platforms. In response to these limitations, research institutes have begun to explore possibilities beyond the confines of the International Mobile Telecommunications-2020 (IMT-2020) framework. As a result, the emergence of 6G technology is anticipated to provide innovative solutions to address the shortcomings of 5G [2].

6G is poised to deliver a high-quality service and enhance energy efficiency, thereby meeting the evolving demands of future applications.

## II. MOBILE NETWORK EVOLUTION

Mobile networks have undergone rapid and remarkable development since their inception in the 1980s. This continuous evolution has given rise to several generations of mobile telephony, with each succeeding generation being introduced approximately every decade [3]. Figure 1 provides a visual representation of the progression of mobile network evolution.

### A. FIRST STEPS FROM 1G TO 3G

The inception of cellular networks dates back to the 1980s when the first generation was introduced, primarily designed for voice services and capable of data rates up to 2.4 kbps. In contrast to the analog systems of the first generation, the second generation embraced digital modulation technologies like Time Division Multiple Access (TDMA) and Code Division Multiple Access (CDMA). This technological shift brought about a significant enhancement, offering data rates of up to 64 kbps and introducing features such as short message service (SMS) [4].

The dawn of the third generation in the year 2000 marked a pivotal moment in mobile network evolution, aiming to provide high-speed data transmission. The 3G network was a game-changer, offering a minimum data transfer rate of 2 Mbps and granting users access to the vast realm of the internet. This transformation enabled an array of advanced services previously unsupported by the 1G and 2G networks, including web browsing, TV streaming, navigation maps, and video services [5].

### B. 4G EVOLUTION

The advent of 4G marked a significant milestone in the evolution of mobile networks, introduced in the late 2000s. This all-IP network was designed to deliver exceptionally high-speed data rates, reaching up to 1 Gbps for downlink and 500 Mbps for uplink transmissions. Its core enhancements included improvements in spectral efficiency and a reduction in latency, making it well-suited to meet the requirements of demanding applications such as digital video broadcasting (DVB), high-definition TV content delivery, and seamless video chat experiences [6]. Notably, Long Term Evolution-Advanced (LTE-A) and Wireless Interoperability for Microwave Access (WiMax) were both recognized as 4G standards. The foundation of 4G, LTE, integrated a blend of existing and novel technologies, including coordinated multiple transmit/receive (CoMP), multiple input multiple output (MIMO), and orthogonal frequency division multiplexing (OFDM) [7].

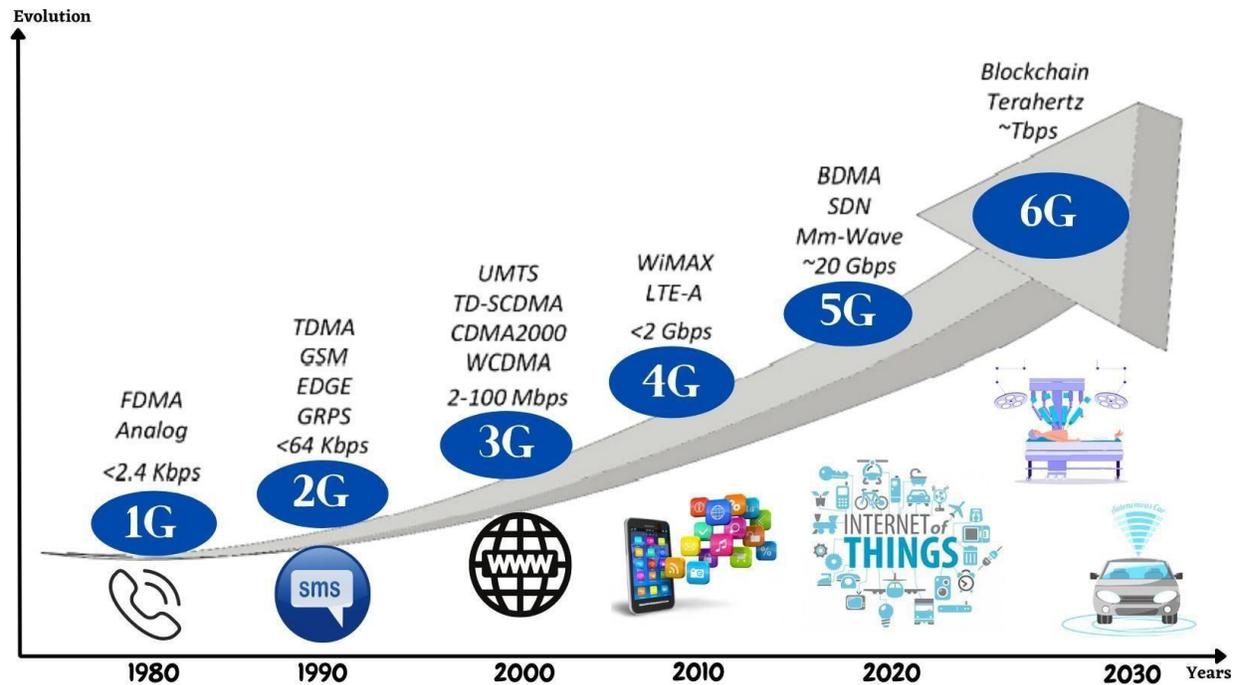

Fig. 1: Evolution of mobile wireless systems.

### C. 5G DYNAMICS

The primary objective of 5G technology is to achieve groundbreaking advancements in terms of throughput, latency, network reliability, and energy efficiency, all while ensuring extensive connectivity. Unlike its predecessors, 5G harnesses not only the new sub-6GHz spectrum, which includes the C-band and the broader 6GHz band but also, for the first time, delves into the millimeter-wave band, enabling significantly higher data rates, reaching up to an impressive ten Gbps. 5G relies on advanced access technologies, including beam division multiple access (BDMA), real-time division multiple access, and filter bank multiple access (FBMC) [1].

To enhance network performance, 5G incorporates a myriad of emerging technologies. These innovations encompass massive MIMO to bolster network capacity, software-defined networks (SDN) for increased network flexibility, device-to-device (D2D) communication to improve spectral efficiency, information-centric networking (ICN) to reduce network traffic, and network slicing to facilitate the swift deployment of numerous services [8].

Within the framework of IMT 2020, three primary usage scenarios have been proposed for 5G technology: enhanced mobile broadband (eMBB) to cater to high-speed data requirements, ultra-reliable low-latency communications (URLLC) for mission-critical applications, and massive Machine Type Communications (mMTC) to support the proliferation of connected devices and the Internet of Things [9].

### III. FUTURE WITH THE 6G

As 5G progresses into its commercial deployment phase, research institutions worldwide are already directing their attention towards the development of 6G, which is slated for deployment around the year 2030. The anticipated capabilities of 6G are poised to usher in a new era of information transmission performance, promising peak data rates of up to one Terabit and achieving ultra-low latency in mere microseconds. A defining feature of 6G is its utilization of communication frequencies in the terahertz range, offering spatial multiplexing capabilities that could potentially provide up to 1000 times the network capacity of 5G systems [10].

One of the central objectives of 6G is the establishment of ubiquitous connectivity, and this is envisaged through the integration of satellite and undersea communication networks, thereby ensuring global coverage [1]. Three distinct categories of 6G services have been outlined to address the evolving needs of users: ubiquitous mobile ultra-broadband (uMUB), ultra-high-speed low-latency communications (uHSLLC), and ultra-high data density (uHDD) [11]. These evolving requirements will be met through the deployment of emerging technologies, including the exploitation of the THz spectrum, federated learning (FL), edge artificial intelligence (AI), compressive sensing (CS), blockchain, and more.

Table 1 serves as a performance comparison tool between 6G and 5G, highlighting the differences and advancements in various key metrics.

### IV. 6G NEW SERVICES

6G services are poised to revolutionize the landscape of mobile communications, reshaping traditional concepts of ultra-reliable low-latency communications (uRLLC), enhanced mobile broadband (eMBB), and massive Machine Type Communications (mMTC). In doing so, they will introduce a host of novel and innovative services, marking a significant departure from the existing paradigm of 5G.

| Performance | 6G | 5G | Effect |
|---|---|---|---|
| Rate | Peak rate: 100 Gbps-1 Tbps | Peak rate: 10 Gbps-20 Gbps | 10–100 times |
| Latency | 0.1 ms, on-time processing | 1 ms | 10 times |
| Traffic density | 100- 10,000 Tbps/square meter | 10 Tbps/square meter | 10–100 times |
| Connection density | Maximum 0.1 billion connections/square meter | 1 million connections/square meter | 100 times |
| Mobility | higher than 1000 km/h | 500 km/h | 2 times |
| Spectrum efficiency | 200–300 bps/Hz | 100 bps/Hz | 2–3 times |
| Positioning | Outdoor 1 meter, Indoor 10 cm | Outdoor 10 meter, Indoor around 1 meter | 10 times |
| Spectrum support | Regular carrier bandwidth 20 Ghz | Sub 6 G Regular carrier bandwidth 100 Ghz, | 50–100 times |
| Reliability | lower than 1/1000,000 | lower than 1/100,000 | 10 times |
| Network efficiency | 200 bits/J | 100 bits/J | 2 times |

TABLE I: Performance comparison between 6G and 5G [4]

### A. MASSIVE URLLC

Smart factories represent one of the key applications of 5G technology, with ultra-reliable low-latency communications (uRLLC) playing a pivotal role in ensuring the reliability and minimal latency of data uplinks within these environments. However, as we look ahead to 6G, there is an anticipation of a significant evolution in the realm of uRLLC. This evolution will introduce an entirely new dimension, giving rise to a novel service known as massive uRLLC (mURLLC). This groundbreaking mURLLC service is not only expected to maintain the stringent requirements of 5G uRLLC but also to coexist seamlessly with the existing massive Machine Type Communications (mMTC) framework [12]. MuRLLC is set to take the delicate balance between reliability, latency, and scalability to unprecedented heights, marking a remarkable advancement in the capability of future communication networks [13].

### B. MOBILE BROADBAND RELIABLE LOW LATENCY COMMUNICATION

The capabilities of enhanced mobile broadband (eMBB) and ultra-reliable low-latency communication (uRLLC) that were sufficient for 5G applications will no longer meet the demands of 6G applications, particularly those involving Extended Reality (XR). These cutting-edge applications necessitate exceptionally high reliability, ultra-low latency, and extremely high data rates. Consequently, a novel service known as Mobile Broadband Reliable Low Latency Communication (MBRLLC) has emerged to address these requirements within 6G networks. MBRLLC is designed to empower 6G networks with the necessary performance attributes to navigate the intricate trade-offs between data rate, reliability, and latency, effectively accommodating the ever-evolving landscape of next-generation applications [14].

### C. MULTI-PURPOSE 3CLS AND ENERGY SERVICES

The forthcoming 6G cellular networks are poised to introduce a spectrum of innovative services, primarily focusing on 3D Connectivity, Localization, and Sensing (3CLS), with the ultimate goal of enabling wireless energy transfer (WET) services for an array of smart devices [15]. Many of these 3CLS applications, encompassing both connectivity and power services (MPS), will place new demands on the 6G network infrastructure. To effectively cater to these requirements, 6G networks must exhibit robust performance attributes, including stringent criteria for latency stability, efficient energy transfer mechanisms, precise localization capabilities, and advanced data processing capabilities. These foundational elements are vital for supporting the diverse range of services and applications expected to define the 6G era [16].

### D. NETWORK SLICING

Network slicing has emerged as a pivotal solution to cater to a diverse range of service requirements within the context of 6G networks. Slicing effectively generates multiple logical networks, each precisely configured to serve a specific type of service, all coexisting on a shared physical infrastructure. Notably, the ITU-R has proposed the implementation of distinct slices for each of the fundamental 5G services, namely ultra-reliable low-latency communications (uRLLC), enhanced mobile broadband (eMBB), and massive Machine Type Communications (mMTC) [17].

To further refine and distinguish the unique requirements of different industries or business areas residing within a given slice, the 3rd Generation Partnership Project (3GPP) specifications have recommended the incorporation of an additional parameter known as a "slice differentiator" (SD). This extension, in a sense, augments the existing horizontal slicing by expanding it vertically. The rationale behind this inclusion lies in the fact that not all industries operating within a single slice can conveniently adopt the same protocol stack. In scenarios where different protocol stacks are necessary, it becomes essential to subdivide the individual slices [18].

However, an aspect that remains underaddressed in current literature is the concept of grouping services that can harmoniously share the same protocol stack within a given slice. This approach carries the potential to unveil the true capabilities of network slicing, enabling it to effectively meet the intricate and multifaceted requirements stipulated in the context of 6G networks.

## V. 6G APPLICATIONS

5G-enabled applications will indeed be at the heart of 6G, even at larger scales (supporting huge, large networks such as safes cities). For 6G the following applications are described:

### A. CONNECTED ROBOTICS AND AUTONOMOUS SYSTEMS (CRAS)

The deployment of new CRAS applications has been the main driver behind the 6G movement. CRAS incorporates flying vehicle delivery systems, autonomous cars, drone swarms, vehicle platoons, and autonomous robotics [19]. These applications are not another IoT uplink service, like many applications introduced in 5G. However, they require stringent latency requirements and extremely high throughput [11].

### B. EXTENDED REALITY

Extended reality (XR) technologies include augmented reality (AR), mixed reality (MR), and virtual reality (VR) [17]. XR is an emerging immersive technology that merges the physical and virtual worlds [11].

Mixed reality is not about overlaying any content from the physical world like AR. However, it eliminates the distinctions between real reality and virtual reality in which computer-generated objects show something obscure in the physical environment.

AR, VR, MR, XR use different sensors to collect data regarding location, orientation and acceleration. This requires strong connectivity, extreme data rates, High resolution and extremely low latency, which should be facilitated by 6G.

### C. INTERNET OF EVERYTHING (IOE)

IoE is an extended version of IoT that includes, things, data, people and processes. The main concept of IoE is to integrate various sensing devices that can be linked to "everything" to identify, monitor the status and make decisions in an intelligent way to create new insights.

IoE sensors are capable of acquiring many parameters such as speed, position, light, biological signals, pressure and temperature. These sensors are used in applications ranging from health systems, safes cities, to various industrial fields. 6G is expected to become a key component of IoE, as it requires the ability to connect N smart devices, where N is scalable and can reach billions. In addition, IoE will need high data rates to easily support N devices with low latency [20].

### D. UAV BASED MOBILITY

UAVs have been widely used for defense and military attack applications, such as remotely operated aircraft, autonomous drones, and others. Over the years, the applications of UAVs have been expanded in the military and civilian fields. For example, drones have been proposed for disaster relief, agricultural plantation protection, traffic monitoring, and environmental sensing [21]. Drones are also expected to be an essential module for future wireless technologies such as 6G, which supports high-speed data transmission for communities living remotely, facing disaster situations such as earthquakes, terrorist attacks, and in the absence of typical cellular infrastructure.

The main characteristics associated with UAVs compared to fixed infrastructure are: ease of deployment, line-of-sight (LoS) connectivity [11] and most importantly, controlled mobility. The rapid development of UAV technologies will enable new areas such as automated logistics and military operations. With the emergence of 6G and IoE, researchers will explore the use of UAV-to-Everything (U2X) network that expands the paradigm of sensing applications by adjusting communication modes to their full potential.

### E. HOLOGRAPHIC TELEPRESENCE

Holographic telepresence (HT) can project realistic three-dimensional (3D) images, in real time and in motion, of remote people and objects, with a high level of realism that rivals physical presence. It can be used for 3D video conferencing in motion, information dissemination etc.

The principle of HT consists of capturing sequences of people and surrounding objects, which are compressed and transmitted over a broadband network in the initial phase. Later, the transmitted information is decompressed on the receiver side and projected with laser beams into the scene to be virtualized. HT technology minimizes the costs of business travel and allows people to appear at many locations simultaneously [22]. In the case of multimedia applications and in order to engage the audience in a full immersive experience, tactile and interactive devices will be implemented using 6G with latency as low as 100 μs with a data rate of several Gbps.

### F. WIRELESS BRAIN-COMPUTER INTERACTIONS (BCI)

In addition to XR applications, smart body implants and BCIs will become essential in 6G to support the healthcare revolution. Current healthcare applications are limited to controlling biometric implants (e.g., controlling prosthetic limbs or controlling functionality with brain implants). In 6G, wireless brain-computer interaction and interface with smart body implants will bring a huge change in the healthcare system. It will introduce new techniques requiring 6G specifications for massive connectivity for innovative operation [23].

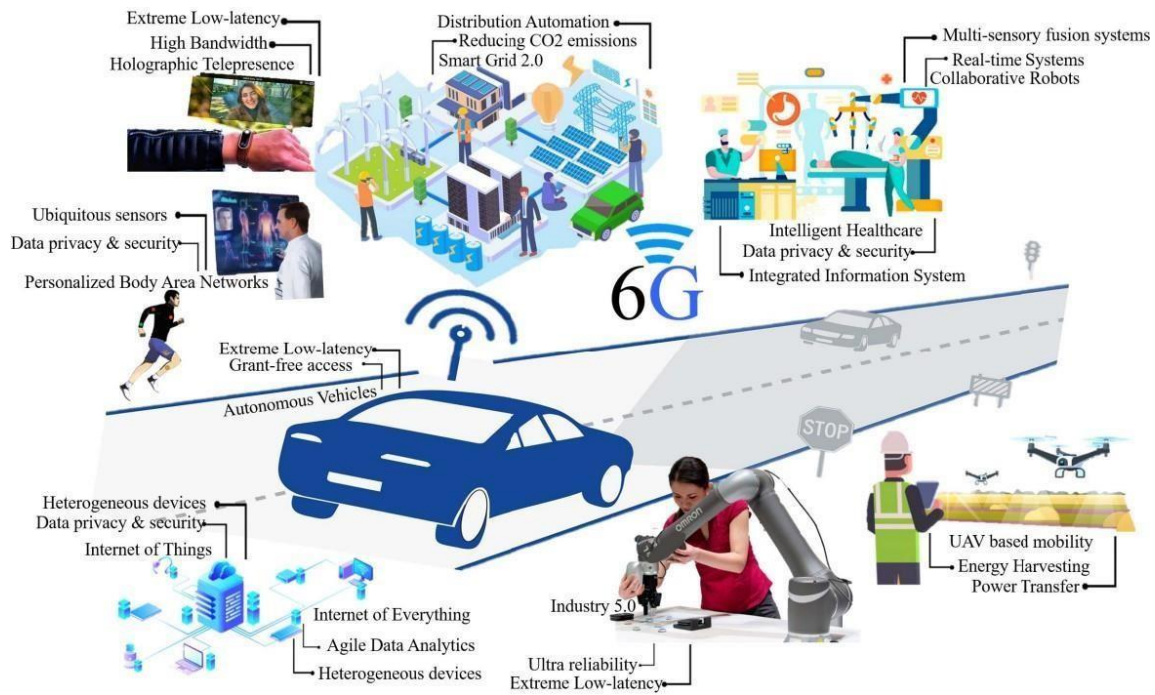

Fig. 2: Some Applications envisaged for 6G.

These applications require stringent requirements (such as high data rates, ultra-low latency and high reliability) like XR. However, BCI applications are much more sensitive than XR, and QoPE must be assured.

### G. PERSONALIZED BODY AREA NETWORKS

Body Area Networks (BANs) with integrated mobile health systems (mHealth) are advancing toward personalized health monitoring and management. These personalized networks can collect health information from multiple sensors, dynamically exchange this information with the environment, and interact with networking services, including social networks [24].

Personalized BAN systems have a wide range of applications, covering both medical and non-medical domains. For example, personalized BANs can be used to avoid the need for wiring in polysomnography tests (also known as sleep disorder diagnosis). Custom BANs have also been used in non-medical applications such as emotion detection, entertainment, and secure authentication applications.

### H. BLOCKCHAIN AND DISTRIBUTED LEDGER TECHNOLOGIES (DLT)

Blockchain and DLT devices are still underdeveloped in many sectors, in addition to being immature. However, they have great potential in these next-generation distributed systems in terms of moving from centralized to distributed systems for the purpose of data validation. They require high connectivity, a synergistic mix of uRLLC and massive machine type communications (mMTC) to maintain low latency, and reliable connectivity [25].

### I. INTELLIGENT HEALTHCARE

The health sector will undergo various evolutions. These will evolve and will now be named healthcare 5.0 with the emergence of digital wellness. AI based smart healthcare will be developed based on several new methodologies including Quality of Life (QoL), Intelligent Wearable Devices (IWD) etc. With recent advances in wearable sensors and computing devices, it is possible to monitor and measure health data in real time. The sensing data collected from wearable devices can be prepared by the nearest edge node and then sent to doctors for remote diagnosis [26]. Moreover, with the realization of holographic communications, touch internet and 6G smart robots, the doctor can perform the surgery remotely. Such telesurgery would eliminate the need for on-site operations and avoid the risks caused by the spread of viruses, especially in the presence of epidemics, such as Coronavirus-19 and other communicable diseases (see fig.3).

### J. INDUSTRY 5.0

Industry 5.0 refers to people working alongside robots and smart machines to add a human touch to the Industry 4.0 pillars of automation and efficiency. The drivers of the future will be based on 6G and IoE as a considerable number of objects in this type of industry will be connected via wires or wireless. The goal will provide various services through the full integration of cloud/edge computing, big data, and AI [27].

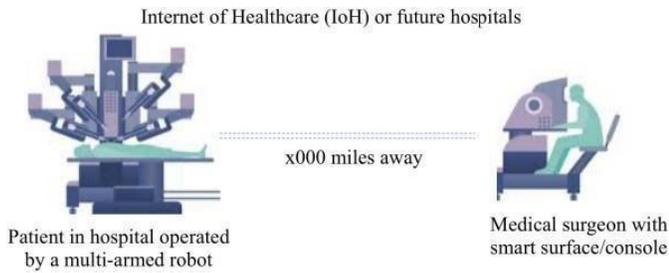

Fig. 3: Application for healthcare in 6G [11].

## VI. CONCLUSION

Mobile networks have evolved very rapidly since their appearance in the 1980s. This article presents the evolution of mobile networks from the 1st generation to the 6th generation, with a focus on the defining principles of this latest generation. What sets 6G apart is its inherent capability to usher in a realm of applications and services that were once beyond the bounds of most people's imagination. This capacity becomes possible because 6G is designed to deliver exceptional quality of service and unrivaled energy efficiency, effectively meeting the diverse demands of future applications.

The 6th Generation envisages to know an unprecedented breakthrough by integrating the traditional terrestrial mobile networks with the emerging space, air and submarine networks in order to provide an access to the network at any time and in any place. What allows to have an underwater connection that will give the possibility to develop several applications, as the underwater tourism, to have direct emissions of the life of the underwater species, to have autonomous submarines and several other applications.